# Bistochastically private release of data streams with zero delay

Nicolas Ruiz

Universitat Rovira i Virgili
Departament d'Enginyeria Informàtica i Matemàtiques
Av. Països Catalans 26, 43007 Tarragona, Catalonia
nicolas.ruiz@urv.cat

**Abstract.** Although the bulk of the research in privacy and statistical disclosure control is designed for static data, more and more data are often collected as continuous streams, and extensions of popular privacy tools and models have been proposed for this scenario. However, most of these proposals require buffers, where incoming individuals are momentarily stored, anonymized, and then released following a delay, thus considering a data stream as a succession of batches while it is by nature continuous. Having a delay unavoidably alters data freshness but also, more critically, inordinately exerts constraints on what can be achieved in terms of protection and information preservation. By considering randomized response, and specifically its recent bistochastic extension, in the context of dynamic data, this paper proposes a protocol for the anonymization of data streams that achieves zero delay while exhibiting formal privacy guarantees. Using a new tool in the privacy literature that introduces the concept of elementary plausible deniability, we show that it is feasible to achieve an atomic processing of individuals entering a stream, instead of proceeding by batches. We illustrate the application of the proposed approach by an empirical example.

**Keywords.** Privacy, anonymization, data streams, randomized response, bistochastic matrices

## 1 Introduction

To manage the tension between data collection and exploratory data analytics on the one hand, and stronger data protection legislation on the other hand, *anonymization* stands out to reconcile both sides, by providing effective privacy protection while offering suitable data utility. Most anonymization techniques and privacy models available today are designed for static data, observed all at once, not revised and published only one time. However, with the accelerating digital transition of societies, a more realistic picture is one of continuous data collections, with large amounts of data incessantly generated. In this paper, we will focus on what is deemed the most challenging type of dynamic data, *data streams* [1], where new records are published continuously, and data freshness is essential.

The problem of providing anonymization solution with formal privacy guarantees for data streams is not new in the literature (e.g. [2,3,4]). However, most of the existing proposals operate with data batches/buffers, where incoming data are first accumulated, anonymized, and then released with a delay. After a thorough review of the literature, we found that proposals based on the currently two most popular privacy models, $k$-anonymity and $\varepsilon$-differential privacy, all make use of data batches to establish privacy guarantees on data streams. While we identify two papers that propose zero-delay, buffer-free procedures [5,6], those are based on perturbative approaches and do not rely on privacy models, and as such do not offer formal privacy guarantees.

Using batches comes unavoidably with a delay, which can be parameterized according to how critical it is to quickly release the freshest data. Thus, batches introduce time as an additional dimension in the classical protection/utility trade-off in privacy. For instance, when a delay has expired and new data needs to be published, one can be forced to release data either highly altered or inadequately protected. Notably, it has been characterized that $k$-anonymity-based methods for data streams tend to enforce privacy guarantees with smaller $k$ and/or higher information loss compared to static data [7]. Also, such methods tend to disregard the alteration in the order of arrival of the data, which can be viewed as an additional utility feature to be preserved in data streams [1].

All in all, the common, and rather intuitive, functioning of the current proposals is that the longer the delay, the larger the batch, and thus the more time one has to suitably perform anonymization, but to the detriment of data freshness. However, in certain application, such as sensor networks, freshness is of the essence, as data can be used in real-time to adjust to anomalous situations. Moreover, using batches and delays entails a huge storage overhead, as it requires doubling the amount of buffer (one for protecting the data that have been stored, in the delay imparted, and one for storing the data that keep entering the stream). Finally, these approaches carry relatively high analytical and computational complexity [7], making the anonymization of data streams a challenging endeavor thus far.

**Contribution and plan of this paper**

Based on this state of affairs, the purpose of this paper is to break ground towards a zero-delay, buffer-free approach for the anonymization of data streams, while providing privacy guarantees. To achieve this, we consider randomized response, a well-established mechanism with rigorous privacy guarantees [8]. To the best of our knowledge, this is the first reported work that considers randomized response in the context of data streams. Furthermore, we will consider the general setting of publishing the data stream *itself*, instead of the publication of count-based or aggregated statistics derived from it (e.g. [9,10]).

Using the recently proposed bistochastic version of randomized response [11], which connects several fields of the privacy literature, we introduce the notion of elementary plausibility, which in turn allows to establish a protocol for a zero-delay publication of data stream with bistochastic privacy guarantees. Those guarantees can moreover be tracked in real time. The remainder of this paper is organized as follows. Section 2 gives some background elements on randomized response and bistochastic privacy, needed later. Section 3 introduces the notion of elementary plausible deniability and then develops a protocol for a zero-delay private publication of data streams. Section 4 presents some empirical results based on this new approach. Conclusions and future research directions are gathered in Section 5.

## 2  Background elements

### 2.1  Randomized response

Let $X$ denotes an original categorical attribute with $1, \ldots, r$ categories, and $Y$ its anonymized version. Given a value $X = u$, randomized response (RR, [8]) computes a value $Y = v$ by using an $r \times r$ Markov transition matrix:

$$P = \begin{pmatrix} p_{11} & \cdots & p_{1r} \\ \vdots & \ddots & \vdots \\ p_{r1} & \cdots & p_{rr} \end{pmatrix} \quad (1)$$

where $p_{uv} = \Pr(Y = v | X = u)$ denotes the probability that the original response $u$ in $X$ is reported as $v$ in $Y$, for $u, v \in \{1, \ldots, r\}$. To be a proper Markov transition matrix, it must hold that $\sum_{v=1}^{r} p_{uv} = 1 \ \forall u = 1, \ldots, r$. $P$ is thus *right stochastic*, meaning that any original category must be spread along the anonymized categories.

The usual setting in RR is that each subject computes her randomized response $Y$ to be reported instead of her true response $X$. This is called the *ex-ante* or local anonymization mode. Nevertheless, it is also possible for a (trusted) data curator to gather the original responses from the subjects and randomize them in a centralized way. This *ex-post* mode corresponds to the Post-Randomization Method (PRAM, [12]). Apart from who performs the anonymization, RR and PRAM operate the same way and make use of the same matrix $P$.

RR is based on an implicit privacy guarantee called *plausible deniability* [13]. It equips the individuals with the ability to deny, with variable strength according to the parameterization of $P$, that they have reported a specific value. In fact, the more similar the probabilities in $P$, the higher the deniability. In the case where the probabilities within each column of $P$ are identical, it can be proved that *perfect secrecy* in the Shannon sense is reached [14]: observing the anonymized attribute $Y$ gives no information at all on the real value $X$. Under such parameterization of $P$, a privacy breach cannot originate from the release of an anonymized data set, as the release does not bring any information that could be used for an attack. However, the price to pay in terms of data utility is high [11].

### 2.2  Bistochastic privacy

We will assume now that the randomized response matrix $P$ above fulfills the additional left stochasticity constraints that $\sum_{u=1}^{r} p_{uv} = 1 \ \forall v = 1, \ldots, r$. This makes $P$ *bistochastic* (left stochasticity implying that any anonymized categories must come from the original categories).

At first sight, one could wonder about the necessity of imposing an additional constraint on RR and its ex-post version PRAM, some well-established approaches for anonymization that have proved their merits over the years. However, it happens that the bistochasticity assumption connects several fields of the privacy literature, including the two most popular models, *k*-anonymity and ε-differential privacy, but also any Statistical Disclosure Control (SDC) method. Indeed, it can be shown that *k*-anonymity guarantees can be reached using a block-diagonal, bistochastic matrix where each block achieves perfect privacy, while ε-differential privacy guarantees can be reached using a circulant, bistochastic matrix. Additionally, it can also be shown that any SDC method can be viewed as a specific case of a more general approach that uses bistochastic matrices to perform anonymization. We refer the reader to [11] for a detailed presentation of these results.

Beyond its unifying properties, the bistochastic version of randomized response offers additional advantages by clarifying and operationalizing the trade-off between protection and utility. Indeed, it is well-known that a bistochastic matrix never decreases uncertainty and is the only class of matrices to do so [15]. Stated otherwise, when a bistochastic $P$ is applied to an original attribute X, its anonymized version Y will always contain more entropy than X. Remark that when $P$ is only right stochastic, as in the traditional approach to RR, no particular relationship emerges.

Now, and as a direct consequence of this last property, the strength of anonymization (and equivalently the strength of plausible deniability), can be measured, as in cryptography for the strength of security, in terms of bits, through $H(P)$, the entropy rate of $P$ (this rate being the average of the entropies of each row of $P$ [16]). In the case of perfect secrecy where all probabilities in $P$ are equal, and that we will denote hereafter by $P^*$, we have $H(P^*) = \log_2 r$, which is the maximum achievable entropy for an $r \times r$ bistochastic matrix. Thus, after the choice of a suitable parameterization, the number of bits that $P$ contains establishes a metric in terms of plausible deniability.

From these results, the definition of bistochastic privacy follows. We provide here the univariate case, where we seek to anonymize only one attribute to prevent disclosure (other definitions at the data set level can be found in [11]):

***Definition 1 (Univariate bistochastic privacy)***: *The anonymized version Y of an original attribute X is $\beta$-bistochastically private for $0 \leq \beta \leq 1$ if:*
    i)        $Y = P'X$ with P bistochastic
    ii)       $\frac{H(P)}{H(P^*)} \geq \beta$.

An anonymized attribute satisfies $\beta$-bistochastic privacy if it is the product of a bistochastic matrix $P$ and the original attribute, and if the entropy rate of $P$ is at least $100\beta\%$ of the maximum achievable entropy. $H(P^*)$ represents the maximum "spending" that can be allocated to privacy, $\log_2 r$ bits. Thus, when $\beta = 1$, all the bits have been spent and the attribute has been infused with the maximum possible amount of uncertainty; in this case, perfect secrecy is achieved and it is clear that $Y = P^{*'}X$ returns the uniform distribution, breaking the link with all other attributes in the dataset and entailing a huge information cost. The other extreme case $\beta = 0$ means that the attribute has been left untouched and no uncertainty has been injected, i.e. $H(P) = 0$. In that case, the data user gets the highest possible utility from the data. As a result, for $0 < \beta < 1$ there lies a continuum of cases where varying amount of uncertainty bits can be injected, which will guarantee a varying amount of protection and information.

Unlike other privacy models, bistochastic privacy makes the trade-off between privacy and information explicit. In fact, by always increasing entropy, it distorts the original information of the data *always* in the same direction, producing anonymized data that are a coarsened version of the original data [11]. Stated otherwise, bistochastically private data are always a compact version of the original data, where some details have been lost. The fact of always coarsening data means that the evaluation of information loss for bistochastically private data is simplified as it can be systematically assessed trough this lens.

Finally, we remark that bistochastic matrices can be applied on both categorical and numerical attributes [11]. In the categorical case, the original proportions of respondents whose values fall in each of the r categories will be changed, which will coarsen the distribution to deliver randomized proportions closer to the uniform distribution. In the numerical case, it will tend to average the numerical values of respondents. In what follows, we will consider first the case of one single numerical attribute.

## 3 Bistochastically private continuous publication of data streams

The approach proposed in this paper makes use of a type of matrix called *T-transforms*. Borrowed from the physics literature [17], to the best of our knowledge they have never been considered before in the context of privacy.

### 3.1 T-transforms and elementary plausible deniability

A T-transform is a linear transformation that acts non-trivially only on two entries of a vector. The matrix representation of a T-transform has the form:

$$T = \lambda I + (1 - \lambda)Q \qquad (2)$$

where $0 \leq \lambda < 1$, $I$ is the identity matrix, and $Q$ is a permutation matrix that just interchanges two coordinates. Remark that *T is a bistochastic matrix*, attaining maximum entropy for $\lambda = 0.5$.

Applied on a numerical attribute $X = (X_1, \ldots, X_r)$, we have:

$$T'X = (X_1, \ldots, X_{j-1}, \lambda X_j + (1-\lambda)X_k, X_{j+1}, \ldots, X_{k-1}, \lambda X_k \\ + (1-\lambda)X_j, X_{k+1}, \ldots, X_r) \qquad (3)$$

In this example, individuals *j* and *k* have been "T-transformed". A casual interpretation of this transformation is that the two individuals have been "mixed", i.e. their attribute's values stayed the same to the extent of $\lambda$, while they inherited the other's value to the extent of $(1 - \lambda)$. While a T-transform is a rather simple tool, it leads to some notable consequences:

***Theorem 1 (Hardy, Littlewood, and Polya [15])***: *The following conditions are equivalent:*
*i) Y= P'X for some bistochastic matrix P.*
*ii) Y can be derived from X by successive applications of a finite number of T-transforms.*

***Theorem 2 (Proschan and Shaked [18])***: *Let $T_1$ be a T-transform of the form (2) with i and k chosen at random from a distribution of r elements that places positive probability on all of the $\binom{r}{2}$ pairs. Assume further that $0 \leq \lambda < 1$ is chosen at random from a distribution with no mass at 0. Repeat this process, using the same distribution for the new pair (i,k) and the same distribution for λ, to obtain $T_2, T_3$ .... Then it holds that:*

$$\lim_{S \to +\infty} \prod_{s=1}^{S} T_s' = P^* \: with \: probability \: 1 \qquad (4)$$

*Theorem 1* states that T-transforms constitute the elementary building blocks of bistochastic matrices. That is, the successive applications of finitely many T-transforms leads to a bistochastic matrix *P*, while *P* can be decomposed as a product of T-transforms. In quantum physics, a T-transform is used to represent the mixture of two particles originally in pure states [17]. In the context of privacy, they can be used to mix two individuals, and in that sense can be interpreted as *elementary plausible deniability*. Indeed, as we saw in RR, the strength of plausible deniability that can be claimed by an individual is conveyed by the parameterization of *P*, and in bistochastic privacy such strength is simply conveyed by the entropy rate of *P*. However, and additionally in the bistochastic case, plausible deniability can be viewed as a succession of elementary steps, where in each one an individual can claim to have been mixed to the extent of $(1 - \lambda)$ with any, and potentially many, other individuals in the data set.

Here lies the fundamental functioning of the approach proposed in this paper: *each individual entering a stream can be equipped with elementary plausible deniability and then released immediately*. The successive application of elementary plausible deniability allows to reconstitute overall plausible deniability, as in RR, which will give the level of privacy guarantee attained by the stream at each point in time. Such approach is feasible only through the bistochastic version of RR, as right stochasticity alone does not allow for a decomposition of the transition matrix through T-transforms [15].

As for *Theorem 2*, it ensures that the repeated application of T-transforms always increases entropy. In fact, the application of infinitely many T-transforms leads to perfect privacy. Such result is rather intuitive in the sense that, by mixing more and more individuals, one is generating a more and more coarsened version of the attribute, to the limit where everyone ends up with the same anonymized value.

It must be noted that the conditions of *Theorem 2* can be relaxed. For instance, it can be shown that by choosing individuals non-randomly and by setting $\lambda = 0.5$, i.e. the maximum possible entropy for a T-transform, one obtains the same result while converging faster to perfect privacy [17].

### 3.2 Data stream anonymization through successive T-transformations

To get a further grasp on the functioning of the approach proposed in this paper, let us consider the hypothetical situation where a numerical attribute observed *all at once* over $r$ individuals is randomized with the bistochastic version of matrix $P$ in (1). Then, an additional individual is added, but the data curator forgets to protect the new value. In that case, necessarily the bistochastic matrix $Q$ over the $r+1$ individuals is of the form:

$$Q = \begin{pmatrix} p_{11} & \cdots & p_{1r} & 0 \\ \vdots & \ddots & \vdots & \vdots \\ p_{r1} & \cdots & p_{rr} & 0 \\ 0 & \cdots & 0 & 1 \end{pmatrix} \quad (5)$$

Over the $r$ individuals, bits have been spent for anonymization (according to the entropy rate of $P$), but no bits have been spent over the $r+1$ new individual, thus the representation of the randomization strength with a one for this individual (having not been randomized, the new individual has a null contribution to the entropy rate of $Q$).

Following the standard formula on the entropy rate of bistochastic matrices [11,16], $H(P) = -\frac{1}{r}\sum_{u,v=1}^{r} p_{uv} \log_2 p_{uv}$ and $H(Q) = -\frac{1}{r+1}\sum_{u,v=1}^{r} p_{uv} \log_2 p_{uv}$, thus $H(Q) = \frac{r}{r+1} H(P)$. In other words, moving from $P$ to $Q$ reduces the entropy rate, which is intuitive in the sense that $Q$ does not make use of the additional possibilities of mixing offered by the new individual.

Now, the data curator realizes her mistake and decides to protect the new individual. However, she would prefer to avoid the cumbersome and time-consuming tasks of creating a new randomization matrix and re-generating a new protected attribute over all the $r+1$ individuals. In that case, a solution is to T-transform the new individual with an already protected one, then updating the protected attribute by adding the protected value of the former and updating the protected value of the latter. Following this procedure, randomization takes place in two steps: first with $P$, then with a T-transform $T$ (of size $r+1$) over two individuals. By associativity of matrix products, there is no need to re-generate entirely the protected attribute. By doing so, the curator reinjects entropy and exploits the new possibilities of mixing. The resulting privacy guarantees are then given by $H(T'Q)$. It can be noted that more than one T-transform can be performed on the new individual (e.g. for 3 T-transforms, the privacy guarantee will be given by $H(T_3'T_2'T_1'Q)$). Given the analytical simplicity of T-transforms, this is a prompt procedure.

The intuition behind the approach proposed in this paper is thus following: a new individual to be anonymized and entering a stream at time $t+1$ can be viewed as an unprotected individual "ghostly" present in the stream in $t$. This means that, when anonymization has ended in $t$ and $r$ individuals have been bistochastically randomized with a matrix $P$ (of size $r\times r$), when the $r+1$ individual enters in $t+1$, $P$ is augmented with a line and a column of zeros, except for the $(r+1)^{th}$ term where it is augmented by a one, as in (5). The $(r+1)^{th}$ individual is then equipped with elementary plausible deniability, being T-transformed with one or more of the $r$ individuals that entered before. The product of the T-transforms and the augmented version of $P$ gives a new bistochastic matrix with an entropy rate expressing the overall privacy guarantee now attained by the stream in *t+1*.

In *t+2*, this process is repeated by augmenting the matrix resulting from the former product and re-doing T-transforms on the $(r+2)^{th}$ individual, and so on. *Theorem 1* ensures that one is building at each stage a bistochastic matrix, while *Theorem 2* ensures that more protection is added along the way. Moreover, and because a T-transform is a basic operation, each new individual can be anonymized and released with zero delay. Compared to other approaches in the literature, here *neither buffer nor delay is required*.

Now, we model a data stream as a discrete time vector processes. An original data stream $\{S_i\}$ is defined as a sequence of continuously incoming tuples $S_i = (I_i, A_i)$ where $I_i$ is the tracking number of the individual *i* to whom $S_i$ corresponds, and $A_i$ is the value of the numerical attribute of that individual. We also denote by $\{S_i^R\}$ the sequence of continuously outputted protected tuples $S_i^R = (I_i, A_i^R)$ with $A_i^R$ the suitably bistochastically randomized version of the numerical attribute. Moreover, we will denote by $P^t$ the bistochastic matrix used to randomized the individuals in the stream before any individuals enter in *t+1*, and by $P^{t(+)}$ its augmented version of the form (5), in *t+1*. The same superscripts will be used for T-transforms.

Furthermore, we will assume that the publication of the stream starts when at least two individuals have already been gathered and randomized with a matrix $P^1$ (thus of size at least 2*)*. We will also assume that the stream has a finite length *L*, albeit it can be as long as necessary. Finally, we remind that, as before, for any parameterization of a given bistochastic matrix *B*, $H(B)$ will denote the entropy rate of *B* while $H(B^*)$ is the maximum possible entropy rate of *B* when all the probabilities within each of its columns are identical (i.e. the perfect privacy case).

The operations for a zero delay, continuous publication of a data stream with bistochastic privacy guarantees, are then performed as follows (for simplicity we consider the case where only one T-transform is performed):

***Algorithm 1***: *Continuous publication of a data stream with bistochastically private guarantees*
**Input at time t**: *A data stream $\{S_i\}$, its protected version $\{S_i^R\}$, a bistochastic matrix $P^t$*
**Output at time t+1**: *A protected data stream $\{S_i^R\}$, a bistochastic matrix $P^{t+1}$, a level of bistochastic privacy $\beta^{t+1}$*
***For*** *t=2,...,L **do***
   1. Read the next tuple $S_i$ entering in *t+1*
   2. Randomly draw $\lambda$ with $0 \leq \lambda < 1$
   3. Randomly select a protected tuple $S_k^R$ who entered $\{S_i\}$ before or a time *t*
   4. T-transform i and k by computing $\lambda A_i + (1-\lambda)A_k^R$ and $\lambda A_k^R + (1-\lambda)A_i$
   5. Compute $P^{t+1} = T^{t+1\prime}P^{t(+)}$ with $T^{t+1}$ the matrix form of the T-transform of step 4, following (2)
   6. Compute $\beta^{t+1} = H(P^{t+1})/H(P^{t+1*})$
***end***
***Return*** $\{S_i^R\}$, $P^{t+1}$, $\beta^{t+1}$

*Algorithm 1* applies a T-transform to any individual *i* entering a stream and then outputs its protected attribute and updates the attribute's protected value of the individual *k* with whom *i* has been T-transformed. Due to the analytical simplicity of a T-transform operation, *i*'s protected attribute, as well as *k*'s updated protected attribute, can be released immediately.

Remark that, as previously mentioned, when anonymizing a data stream the order of arrival of tuples is an additional information to be preserved compared to static data sets: the less reordering, the more utility [1]. In previous contributions, the use of buffers and delay constraints altered the ordering, but such effect was often disregarded (e.g. [2]), or additional operations were necessary for trying to mitigate it (e.g. [19]). In this paper, the tuples' order is automatically preserved, as *Algorithm 1* is not buffering them.

*Algorithm 1* outputs levels of bistochastic privacy $\beta^{t+1}$, allowing the data curator to monitor the stream's degree of protection in real time, which is an appealing feature. However, as it stands it is mute about how to control and adjust this degree. This can be introduced easily in two ways. First, following the discussion above, $\lambda$ does not necessarily need to be random. One can set $\lambda=0.5$ and use the maximum possible entropy in each T-transform. Second, nothing precludes to perform more than one T-transform on each new individual entering the stream. This will necessarily introduce more protection following *Theorem 2,* but it will not have an impact on the delay to release the new protected tuple, given that several T-transforms can be performed quickly. By using these two levers, *a data curator can tune up or down the level of protection in the stream,* which is also an appealing feature. All that it requires is a slight and straightforward generalization of *Algorithm 1* along the lines just exposed, that we will not write here due to space constraints.

### 3.3   Application to categorical attributes

We described so far the case for a single numerical attribute. We now turn to the case of a categorical one and first remark that, in practice, the number of categories in such attributes is unlikely to change often, if at all, as long as a stream progresses. While numerical attributes like online purchases will be different each time for each individual, the categorizations of variables like professional occupation, region of residence… are less likely to change. In that case, the use of randomized response, and its bistochastic version, is identical on a data stream and a static data set: each individual entering the stream will be randomized according to a matrix *P* parameterized with some transition probabilities over a pre-defined number of *r* categories at the beginning of the stream, and *Algorithm 1* is not required here. New individuals will be able to plausibly deny a reported value according to *P* and its associated entropy rate, in the same way individuals already incorporated in the stream can.

However, should a given categorization happens to change along the stream, *Algorithm 1* can be used. If in *t* an individual enters the stream with a previously unforeseen category that needs now to be taken into account for protection, a T-transform will mix the new category with a previous category. Hence, assuming that in *t* the categorical variable has *r* categories, of which *j*, and in *t+1* the *r+1* new category is T-transformed with $j$, $\lambda$ in the T-transform is now interpreted as the probability that the individual entering the stream, and reporting as true value the category $r+1$, will be reported in the protected stream also as $r+1$, and reported as $j$ with a $(1 - \lambda)$ probability. Moreover, all individuals in the $j$ category already in the protected stream are randomized a second time to inherit, or not, the category $r+1$, and their category is updated in the protected stream. Thus, *Algorithm 1* applied on a categorical variable operates in the same way than on a numerical variable, except that *it mixes categories instead of individuals*.

Of course, only mixing a new category with a former one could appear insufficient for protection, but here again several T-transforms can be performed for additional category mixing. The result of this process is a bistochastic matrix $P^{t+1}$, and its entropy rate gives the new privacy guarantees over the $r+1$ categories. This matrix can now be used to randomize the next entering individuals until a new category emerges, which will require to repeat the procedure.

### 3.4 Application to several attributes

We now consider the case of several attributes, assuming that they are all dynamic in the stream, i.e. individuals keep entering with values in all attributes, categorical of numerical. We consider a stream with *M* attributes, and the continuously incoming tuples are now denoted by $S_i = (I_i, A_{1,i}, \ldots, A_{M,i})$. In that context, following [11] the characterization of privacy guarantees at the data stream level can be twofold:

- Each attribute is dealt with separately, and the privacy guarantee in $t+1$ is $\beta^{t+1} = \frac{\sum_{m=1}^{M} H_m(P^{t+1})}{\sum_{m=1}^{M} H_m(P^{t+1*})}$, where $H_m(.)$ denotes the entropy rate of the bistochastic matrix applied on attribute $m$. This case has the merit of simplifying the implementation of bistochastic privacy on a whole data stream. However, the drawback is that, because entropy is sub additive, one injects more bits than in the case of dealing directly with the joint distribution. More protection is applied and, as a result, more information is lost. In particular, the dependencies between attributes may end up getting more degraded than necessary (except in the case where all the original attributes are independent).

- A way to avoid information loss when attributes are dependent is to apply a bistochastic matrix directly on the joint distribution $X_J = X_1 \times \ldots \times X_M$. In that case, the privacy guarantee will be given by $\beta^{t+1} = \frac{H(P_J^{t+1})}{H(P_J^{t+1*})}$, where $P_J^{t+1}$ denotes the bistochastic matrix applied directly on the joint distribution of the stream at time $t+1$. While such approach appears in principle the most appropriate one, its computational cost may however result in practical hurdles. Indeed, $P_J^{t+1}$ may reach a very large size as the stream progresses, not least if it contains many numerical attributes. However, one can note that, in the product $T^{t+1'}.P^{t(+)}$ of *Algorithm 1*, $T^{t+1}$ is a sparse matrix that acts only on two lines of $P^{t(+)}$. That means that, as the stream grows, it is not computationally unrealistic to repeat *Algorithm 1* on a joint distri-bution and computing the level of privacy guarantees. Still, like other privacy models, bistochastic privacy is not immune to the curse of dimensionality [20].

## 4 Empirical illustrations

We start by noting that, to achieve bistochastic privacy, one just needs to parameterize bistochastic matrices, as one does not need to observe the actual data [11]. Therefore an agent, independent of the data controller, say a "data protector", can generate the appropriate matrices. The associated parameter $\beta$ for those matrices will depend on the environment and the desired protection-utility trade-off. As a result, performing bistochastic privacy can be viewed as close to performing cryptography, where an encryption key is generated independently of the message to be protected.

In the example below, we act as a data protector who generates bistochastic matrices along a stream that ends at $r=100$ individuals, and where each one enters the stream with one value for a numerical attribute. We initiate the stream with two individuals who have been randomized with the bistochastic matrix $\begin{pmatrix} 0.8 & 0.2 \\ 0.2 & 0.8 \end{pmatrix}$ (thus with $\beta = 72\%$). The stream then starts at $r=3$. The data protector will use *Algorithm 1* to release individuals immediately, albeit under different settings: *i)* Algorithm 1 as formulated, where only one T-transform is performed on each new individual and $\lambda$ is randomly generated; *ii)* only one T-transform is performed but $\lambda$ is always set at 0.5; *iii)* a random number of T-transforms between 2 and 10 are performed on each new individual and $\lambda$ remains set at 0.5. Those settings are set to go each time in the direction of more entropy and thus higher $\beta$'s, as the stream progresses. The results of this exercise are shown in Table 1.

**Table 1.** Example of bistochastic guarantees on a data stream.

| Number of individuals having entered the stream | Bistochastically private guarantees (β's) under different settings of Algorithm 1 | | |
|---|---|---|---|
| | i) | ii) | iii) |
| 20 | 40% | 52% | 69% |
| 40 | 38% | 49% | 68% |
| 60 | 35% | 50% | 72% |
| 80 | 31% | 50% | 74% |
| 100 | 30% | 48% | 75% |

In this example, under *i)* privacy guarantees are continuously degraded, as only one T-transform is performed on each new individual entering the stream. Thus, even if each one can claim elementary plausible deniability, the overall privacy guarantees of the stream are comparatively low and decreasing, as the mixing opportunities offered by the new individuals are only partially exploited. In comparison, under *ii)* using T-transforms with maximum entropy delivers stricter guarantees at each stage. While new individuals are only mixed one time with another, they are so with maximum entropy each time. However, privacy guarantees are still decreasing. Finally, under *iii)* privacy guarantees are stronger overall and continuously increasing, as performing more than one T-transform at each stage allows to exploit to a larger extent the mixing possibilities offered by new individuals. Thus, as mentioned above *Algorithm 1* can be tuned in order to adjust for the level of privacy guarantees one wishes to achieve, by performing a varying number of T-transforms of varying entropy.

To conclude this section, we remark that the existence of a generic rule to devise the increase in entropy after each T-transform is not known in the physics literature. Indeed, what has only been characterized so far is that such process is sub-additive in nature, following a general result below on the entropy rate of the product of two bistochastic matrices:

***Theorem 3 (Zyczkowski, Kus, Slomczynski, and Summers [21]):*** *Let A and B be two bistochastic matrices. Then it always holds that:*
$$\max\{H(A), H(B)\} \leq H(AB) \leq H(A) + H(B)$$

## 5  Conclusions and future work

This paper considered bistochastic privacy, a recently proposed type of randomized response, in the context of data streams anonymization. By introducing in privacy a specific class of bistochastic matrices called T-transforms, and which underlies a new notion of elementary plausible deniability that we also introduced, we derived a zero delay, buffer-free protocol with rigorous privacy guarantees for the private release of data streams. Our proposal for such data environment reestablishes the original protection/utility trade-off in privacy (as in the case when protecting static data sets), in comparison with the current state of the literature for dynamic data, where time is required as a third dimension to consider. Moreover, and due to the properties of bistochastic privacy, the appraisal of this trade-off is clarified in terms of bits (again as for static data). Furthermore, this trade-off can be tracked in real time as a stream progresses.

This paper opens several lines for future research. One of them is to conduct further empirical work on real-life streams. Another is to identify which structures and sequence of T-Transforms can be conducive to bistochastic matrices with differential private or *k*-anonymous privacy guarantees, by leveraging the unifying properties of bistochastic privacy. Yet another challenge is to devise a formal rule that can account for the changes in bits as T-transforms are successfully applied, in order to ease further the control of protection by the data curator. As we mentioned, few is currently known on this issue [21,22]. Finally, an investigation of the compatibility, within the framework proposed in this paper, of the recent solutions to mitigate the dimensionality problem in RR [20], is warranted.

## References

1. J. Parra-Arnau, T. Strufe, J. Domingo-Ferrer, "Differentially private publication of database streams via hybrid video coding", *Knowledge-Based Systems*, Vol. 247, 108778, Jul 2022.

2. J. Cao, B. Carminati, E. Ferrari, K. Tan, "CASTLE: Continuously anonymizing data streams", *IEEE Trans. Depend. Secure Comput*. 99 (2009).
3. A. Otgonbayar, Z. Pervez, K. Dahal, S. Eager, "K-VARP: K-anonymity for varied data streams via partitioning", *Information Sciences*, Vol. 467, 2018, Pages 238-255.
4. K. Guo, Q. Zhang, "Fast clustering-based anonymization approaches with time constraints for data streams", *Knowledge-Based Systems*, 46 (2013): 95-108.
5. N. Jha, T. Favale, L. Vassio, M. Trevisan, M. Mellia. "z-anonymity: Zero-Delay anonymization for data streams", *2020 IEEE International Conference on Big Data*, pp. 3996-4005.
6. S. Kim, M.K. Sung, Y.D. Chung, "A framework to preserve the privacy of electronic health data streams", *J. Biomed. Inform.*, 50 (2014), 95–106.
7. P. Wang, J. Lu, L. Zhao, J. Yang, "B-castle: an efficient publishing algorithm for k-anonymizing data streams", *2010 Second WRI Global Congress on Intelligent Systems*, 2010, Vol. 2, pp. 132-136, IEEE.
8. A. Chaudhuri, R. Mukerjee, *Randomized Response : Theory and Techniques*. Marcel Deker, 1988.
9. L. Fan, L. Xiong, "An adaptive approach to real-time aggregate monitoring with differential privacy", *IEEE Trans. Knowl. Data Eng.*, 2013, 26(9), 2094-2106.
10. F. Fioretto, P. Van Hentenryck, "Optstream: releasing time series privately", *Journal of Artificial Intelligence Research*, 2019, 65, 423-456.
11. N. Ruiz, J. Domingo-Ferrer, "Bistochastic privacy", *Modeling Decisions for Artificial Intelligence-MDAI2022, Sant Cugat, Espanya, In Lecture Notes in Artificial Intelligence*, vol. 13408, pp. 53-67.
12. PP. de Wolf, "Risk, utility and PRAM", *Privacy in Statistical Databases. PSD 2006, in Lecture Notes in Computer Science*, vol 4302, pp. 189-204.
13. J. Domingo-Ferrer, J. Soria-Comas, "Connecting randomized response, post-randomization, differential privacy and t-closeness via deniability and permutation", 2018, https://arxiv.org/abs/1803.02139.
14. C.E. Shannon, "Communication theory of secrecy systems", *Bell System Technical Journal* 28, 4 (1949), 656–715.
15. A. W. Marshall, I. Olkin, B. Arnold, *Inequalities: theory of majorization and its applications*, Springer Series in Statistics, 2011.
16. K. Jacobs, *Quantum Measurement Theory and its Applications*, Cambridge University Press (2014).
17. I. Bengtsson, A. Ericsson, "How to mix a density matrix", *Phys. Rev. A*, 2003, 67, 012107.
18. F. Proschan, and M. Shaked, "Random averaging of vector elements". *SIAM J. Appl. Math.*, 1984, 44, 587–590.
19. J. Domingo-Ferrer, J. Soria-Comas, R. Mulero-Vellido, "Steered microaggregation as a unified primitive to anonymize data sets and data streams", *IEEE Transactions on Information Forensics and Security*, Vol. 14, no. 12, pp. 3298-3311, 2019.
20. J. Domingo-Ferrer, J. Soria-Comas, "Multi-dimensional randomized response", *IEEE Trans. Knowl. Data Eng.*, Vol. 34, no. 10, pp. 4933-4946, 2022.
21. K. Zyczkowski, M. Kus, W. Slomczynski, H. Sommers, "Random unistochastic matrices", *J. Phys. A: Math. Gen.*, 2003, 36, 3425.
22. K.T. Arasu, Manil T. Mohan, "Entropy of orthogonal matrices and minimum distance orthostochastic matrices from the uniform van der Waerden matrices", Discrete Optimization, 2019, Vol. 31, pp. 115-144.